\documentclass[a4paper,12pt,final]{amsart} 
\allowdisplaybreaks[1]
\usepackage{amssymb,amsthm}
\usepackage[TS1,T1]{fontenc}    
\RequirePackage{ifthen}
\usepackage{paralist}
\usepackage[linktocpage,pdftex,final]{hyperref}
\RequirePackage{xcolor}

\usepackage[capitalise]{cleveref}
\Crefname{figure}{Figure}{Figures}
\usepackage{tikz}
\usepackage{aliascnt}           
\newcommand{\mynewtheorem}[4][]{
  \ifthenelse{\equal{#1}{}}{   
    \newtheorem{#2}{#3}        
  }{
    \newaliascnt{#2}{#1}       
    \newtheorem{#2}[#2]{#3}    
    \aliascntresetthe{#2}      
  }
  \crefname{#2}{#3}{#4}        
}
\newtheorem*{theorem*}{Theorem}
\newtheorem*{model*}{Model}
\newtheorem*{thirring*}{Thirring's inequality}
\theoremstyle{remark}
\newtheorem{remark}{Remark}

\mynewtheorem{thm}{Theorem}{Theorems}
\mynewtheorem[thm]{model}{Model}{Models}
\mynewtheorem[thm]{prop}{Proposition}{Propositions}
\mynewtheorem[thm]{cor}{Corollary}{Corollaries}
\mynewtheorem[thm]{lemma}{Lemma}{Lemmas}
\mynewtheorem[thm]{example}{Example}{Examples}
\mynewtheorem[thm]{definition}{Definition}{Definitions}

\providecommand{\esssup}[1][]{\operatorname*{\ifthenelse{\equal{}{#1}}{}{\mathit{#1}-}ess\,sup}}
\newcommand{\supp}{\operatorname{supp}}

\newcommand{\norm}[2][]{\lVert#2\ifthenelse{\equal{}{#1}}{\rVert}{\mathclose{\rVert}_{#1}}}
\newcommand{\bignorm}[2][]%
  {\bigl\lVert#2\bigr\rVert\ifthenelse{\equal{}{#1}}{}{_{#1}}}
\newcommand{\Bignorm}[2][]%
  {\Bigl\lVert#2\Bigr\rVert\ifthenelse{\equal{}{#1}}{}{_{#1}}}
\newcommand{\biggnorm}[2][]%
  {\biggl\lVert#2\biggr\rVert\ifthenelse{\equal{}{#1}}{}{_{#1}}}
\newcommand{\Norm}[2][]%
  {\left\lVert#2\right\rVert\ifthenelse{\equal{}{#1}}{}{_{#1}}}
\newcommand{\spr}[2]{\langle#1,#2\rangle}
\newcommand{\Id}[1][]{\operatorname{Id\ifthenelse{\equal{}{#1}}{}{_{#1}}}}
\newcommand{\Tr}{\operatorname{Tr}}
\usepackage{bbm}
\newcommand{\E}{\mathbbm E}
\newcommand{\Prob}{\mathbbm P}

\newcommand{\bN}{\mathbbm N}

\newcommand{\bR}{\mathbbm R}
\newcommand{\bZ}{\mathbbm Z}

\newcommand{\cD}{\mathcal D}
\newcommand{\cH}{\mathcal H}

\newcommand{\qtextq}[1]{\quad\text{#1}\quad}

\let\originald\d 
\renewcommand{\d}{\ifthenelse{\boolean{mmode}}{\mathrm d}{\originald}}
\def\Int#1d{\int#1\,\d} 

\providecommand{\abs}[2][]{{\lvert#2\rvert}\ifthenelse{\equal{#1}{}}{}{_{#1}}}
\providecommand{\bigabs}[2][]{{\bigl\lvert#2\bigr\rvert}\ifthenelse{\equal{#1}{}}{}{_{#1}}}

\providecommand{\bigfloor}[1]{{\bigl\lfloor#1\bigr\rfloor}}

\providecommand{\setsize}[1]{\operatorname{\#}#1}

\providecommand{\om}{\omega}

\providecommand{\vol}{\abs}

\providecommand{\ifu}[1]{\chi\ifthenelse{\equal{#1}{}}{}{_{#1}}}
\providecommand{\Vper}{V_{\mathrm{per}}}
\providecommand{\Hper}{H_{\mathrm{per}}}

\providecommand{\Ico}[2]{\mathopen[#1,#2\mathclose)}
\providecommand{\Ioc}[2]{\mathopen(#1,#2\mathclose]}

\providecommand{\Laplace}{\Delta}

\providecommand{\from}{\colon}

\providecommand{\argmt}{\mathchoice{{}\cdot{}}{{}\cdot{}}{{}\bullet{}}{{}\bullet{}}}
\providecommand{\dx}{\d x}
\providecommand{\molly}[1]{\tilde\chi\ifthenelse{\equal{#1}{}}{}{_{#1}}}%
\newcounter{todo}
0
\RequirePackage{xcolor}

\makeatletter 
\newif\ifdr@ft\dr@ftfalse
\relax
\newcounter{const@ntNo}
\newcommand{\reloadConst@nt}{%
  \stepcounter{const@ntNo}%
  \edef\genericConst@ntInternal{C_{\theconst@ntNo}}%
}
\newcommand{\genericConstant}[1]{%
  \reloadConst@nt%
  \expandafter\let\csname #1\endcsname\genericConst@ntInternal%
  \ifthenelse{\boolean{dr@ft}}{%
    \marginpar{\textcolor{gray}{\fbox{#1 = $\csname #1\endcsname$}}}%
  }{}%
}
\makeatother

\begin{document}

\title[Lifshitz asymptotics for breather models]
  {Lifshitz tails for Schr\"odinger operators with random breather potential}
\author{Christoph Schumacher \and Ivan Veseli\'c }
\address{TU Dortmund, Fakult\"at f\"ur Mathematik, 44227 Dortmund\\ Germany}

\begin{abstract}
  We prove a Lifshitz tail bound on the integrated density of states
  of random breather Schr\"odinger operators.
  The potential is composed of translated single site potentials.
  The single site potential is an indicator function of set~$tA$ where~$t$
  is from the unit interval and~$A$
  is a measurable set contained in the unit cell.
  The challenges of this model are:
  Since~$A$ is not assumed to be star-shaped
  the dependence of the potential on the parameter~$t$ is not monotone.
  It is also non-linear and not differentiable.\\[1em]

Nous prouvons une in\'egalit\'e de Lifchits pour la densit\'e d'\'etats int\'egr\'ee
pour des op\'erateurs de Schr\"odinger avec potentiel al\'eatoire de breather.
Plus pr\'ecis\'ement, le potentiel est compos\'e de translations d'un potentiel
simple site, qui est une fonction caract\'eristique de l'ensemble $t A$, o\`u
$t\in [0,1]$ et $A\subset[-1/2,1/2]^d$ est mesurable.
L'enjeu de ce mod\`ele r\'eside dans les propri\'et\'es suivantes:
Puisque nous n'assumons pas que la partie $A$ soit \'etoil\'ee, le potentiel
est une fonction non monotone de la variable $t$.
De plus, la d\'ependance est non lin\'eaire et non diff\'erentiable.
\end{abstract}
\maketitle

\section{Model and result}
We prove a Lifshitz tail bound on the integrated density of states (IDS)
for a random Schr\"odinger operator with breather potential.
In comparison to other models,
in particular the  well studied alloy type potential,
the major challenge in our model is that it is neither monotone nor linear
as a function of the random parameter(s).
This is a feature shared with the random displacement model \cite{KloppLNS}
and with random quantum waveguides \cite{BorisovV-13},
to name just two problems which have been studied recently in the literature.
Moreover, the operator family under consideration here is not analytic
in the sense of Kato. In fact, its derivative does not exist as a bounded operator.
This is due to the fact that the most natural single site potential
is a characteristic function of a measurable set.

The direct predecessor of our work is \cite{KirschV-10}.
Below we will compare the results of \cite{KirschV-10} with ours.

\begin{model*}
Let $A\subset\cD :=[-1/2,1/2]^d\subset\bR^d$ be measurable
with Lebesgue measure $\vol A\in\Ioc0{1/2}$,
$tA:=\{tx\in\bR^d\mid x\in A\}$,
$u(t,x):=\ifu{tA}(x)$ the indicator function,
and $\lambda_j\from\Omega\to[0,1]$, $j\in\bZ^d$,
an i.\,i.\,d.\ sequence of random variables satisfying
\begin{equation*}
  \Prob(\lambda_0=0)<1\text,\qquad
  \forall\varepsilon>0\colon
  \Prob(\lambda_0\in[0,\varepsilon])>0\text.
\end{equation*}
For $\Vper\in L^\infty(\bR^d)$ periodic with respect to~$\bZ^d$,
we define the unperturbed background operator
\begin{equation*}
  \Hper:=-\Laplace+\Vper
  \qquad\text{with domain $W^{2,2}(\bR^d)$}
\end{equation*}
and its random perturbation
\begin{equation}\label{eq:operator}
  \begin{aligned}
    H_\omega&:=\Hper+W_\omega
    :=\Hper+\sum_{j\in\bZ^d}u(\lambda_j(\omega),\argmt-j)
    \\&\phantom:
    =\Hper+\sum_{j\in\bZ^d}\ifu{\lambda_j(\omega)A}(\argmt-j)
  \end{aligned}
\end{equation}
\end{model*}

A Borel-Cantelli argument shows that
\begin{equation*}
  E_0:=\inf\sigma(\Hper) =\inf\sigma(H_\omega)    \qquad\text{a.\,s.}
\end{equation*}
Consequently, the IDS
\begin{equation*}
  N\from\bR\to[0,\infty)
  \text,\quad
  N(E):=\E[\Tr\chi_\cD\chi_{(-\infty,E]}(H_\omega)]
\end{equation*}
vanishes below~$E_0$ and is positive above~$E_0$.

For Schr\"odinger operators with ``truly'' random potential
one expects that~$N$ is very thin near~$E_0$.
In fact, we prove for the above model

\genericConstant{Cfront}%
\genericConstant{Cexp}%
\begin{theorem*}
  There exist $\Cfront,\Cexp\in(0,\infty)$ and $E'>E_0$
  such that for all $E\in(E_0,E']$
  \begin{equation}\label{eq:lifshitz}
    N(E)
    \le \Cfront(E-E_0)^{d/2}\exp(-\Cexp(E-E_0)^{-d/2})\text.
  \end{equation}
\end{theorem*}

\begin{figure}
  \begin{tikzpicture}
    [baseline=0mm
    ,mylabel/.style={fill=white,inner xsep=.1mm,inner ysep=.5mm}
    ]
    \draw [gray] (0,0) -- (5,-2.5);
    \draw [gray] (0,0) -- (4,2);
    \draw [fill=black] circle (.3mm)
      node [label=above:$0$] {}
      node [mylabel] at (-0.1,-0.4) {$\supp u_0$};
    \foreach \l in {1,{2/3},{1/3}} {
      \draw [fill=gray] (5*\l,-2.5*\l)
        .. controls (8*\l,-2.5*\l) and (6*\l,2*\l)  .. (4*\l,2*\l)
        .. controls (6*\l,1*\l)    and (4*\l,-2*\l) .. (5*\l,-2.5*\l);
      \draw node [mylabel] at (5*\l-0.1,-2.5*\l-0.4) {$\supp u_{\l}$};
      }
    \draw node at (5.6,-1.1) {$A$};
  \end{tikzpicture}
  \caption{Support of single site potential $u_t$
    for different values of $t$  with arbitrary base set~$A$}
  \label{fig:genbreather}
\end{figure}
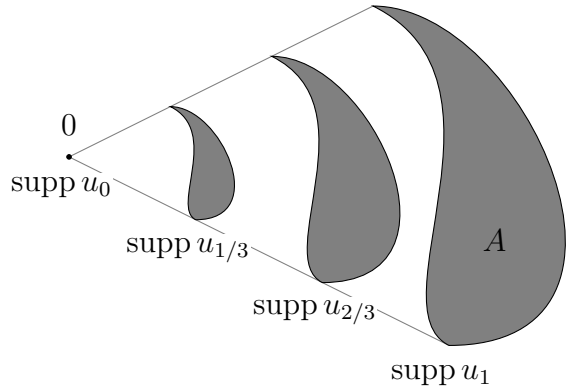

\begin{remark}
In a longer companion paper \cite{SchumacherV-long} we will discuss more details, in particular:
\begin{itemize}
\item more general breather models than~\eqref{eq:operator},
  in fact an abstract non-linear model incorporating the usual breather
  and alloy type models,
\item a lower bound complementary to~\eqref{eq:lifshitz},
\item applications, in particular initial length scale estimates and its
  team work with recent Wegner estimates \cite{NakicTTV-15,TaeuferV-15,  NakicTTV-16} to yield Anderson localization,
\item the history of the problem and previous literature.
\end{itemize}
In contrast, in the present paper we want to keep the presentation simple
and concentrate on the main idea of our proof for a very intuitive model.
\end{remark}
\begin{remark}
  We compare our result to its direct predecessor in \cite{KirschV-10}.
  This is also the easiest way to point out the differences in the two proofs.

  In \cite{KirschV-10}, a Lipschitz or differentiability condition was required
  for the single site potential, namely
  \begin{equation}\label{eq:diffble}
    \frac{\d}{\d\lambda}u(\lambda,\argmt)\in L^\infty(\bR^d)\text.
  \end{equation}
  For our choice $u(\lambda,x)=\ifu{\lambda A}(x)$, the derivative
  $\frac{\d}{\d\lambda}u(\lambda,\argmt)$ is not even a function,
  let alone an element of~$L^\infty$.
  Condition~\eqref{eq:diffble} was used in \cite{KirschV-10}
  to linearize the non-linear model and apply Temple's inequality in a similar
  fashion as in the case of the linear alloy type model.
  Furthermore, in \cite{KirschV-10} it is assumed that $\lambda\mapsto u(\lambda,x)$
  is isotone for almost every $x\in\bR^d$.
  This is obviously \emph{not the case} for $\lambda\mapsto\ifu{\lambda A}$
  unless~$A$ is star shaped with center~$0$,
  a condition we \emph{do not} impose.
  In our proof, we use merely monotonicity-on-average,
  roughly speaking the fact that
  \begin{equation*}
    \int\ifu{t A}\,\dx=\vol{tA}
  \end{equation*}
  is increasing in~$t$.
  Finally,
  let us stress that we do not assume any topological properties of~$A$,
  neither openness nor regularity of the boundary, see \cref{fig:genbreather}.
  In particular, $A$ may be a fractal set.

To avoid the assumptions which have been necessary in \cite{KirschV-10},
we do not use Temple's inequality, but Thirring's inequality \cite{Thirring-94}
instead.
Note that Thirring's inequality was used in the pioneering work
\cite{KirschM-83a}
on Lifshitz Tails for random Sch\"odinger operators of alloy-type,
but has been abandoned in favor of Temple's inequality in subsequent
papers starting with \cite{Simon-85b}.
\end{remark}

\section{Proof}
The proof of the Theorem 
relies on the following perturbation bound,
whose proof via the projection method can be found in \cite{Thirring-94},
see the forthcoming \cite{SchumacherV-long} for details.
\begin{thirring*}
  Let~$H$ be a self-adjoint operator on a Hilbert space~$\cH$,
  such that $E_1(H):=\min\sigma(H)$
  is a simple eigenvalue with normalized eigenstate~$\psi\in\cH$ and 
  $E_2(H):=\inf(\sigma(H)\setminus\{E_1(H)\})>E_1(H)$.
  Let~$V$ be an invertible, positive operator on~$\cH$,
  such that   $\spr\psi{V^{-1}\psi}>0$.
  Then
  \begin{equation*}
    \min\{E_1(H)+\spr\psi{V^{-1}\psi}^{-1},E_2(H)\}
      \le E_1(H+V)\text.
  \end{equation*}
\end{thirring*}

\genericConstant{Cgap}%
\genericConstant{Cld}%
\begin{proof}[Proof of the Theorem
]
  We define the relevant index set $I_L:=\Ico{-L}L^d\cap\bZ^d$ for $L\in\bN$
  and assemble the box $\Lambda_L:=I_L+\cD$. 
  With~$\vol{\argmt}$ for Lebesgue measure and $\setsize{}$
  for cardinality we have $\vol{\Lambda_L}=\setsize{I_L}$.
 We need to show \eqref{eq:lifshitz} only for points~$E$  of continuity of~$N$,
 since~$N$ is monotone and the right hand side is continuous in~$E$.

Let $H_\omega^L$ denote  the restriction of~$H_\omega$ to~$\Lambda_L$   with Neuman boundary conditions.
  Weyl's bound implies that for non-negative single site potentials 
  \begin{equation*}
    N(E)\le\Cfront E^{d/2}\Prob\{\om\mid E_1(H_\omega^L)\le E\}
  \end{equation*}
  for $L\in\bN$ and points $E\in\bR$ of continuity of~$N$.
  Here we will for simplicity assume that $\Vper=0$, in particular $E_0=0$.
  (For  $\Vper\ne0$,  one needs to use in the following arguments Mezincescu boundary conditions \cite{Mezincescu-87} instead,
  as we elaborate in detail in \cite{SchumacherV-long}.)
  It is thus sufficient to derive an exponential bound on the probability
  that the first eigenvalue~$E_1(H_\omega^L)$ of~$H_\omega^L$
  does not exceed~$E$ for a suitably chosen~$L=L_E$.

  In order to apply Thirring's inequality,
  we need the random potential to be strictly positive.
  We therefore regularize the potential by letting
  \begin{equation*}
    H_0^L:=-\Delta^L+-\gamma_L\quad\text{and}\quad
    V_\omega:=W_\omega+\gamma_L
  \end{equation*}
  with $\gamma_L:=\Cgap/(2L^2)$ and $\Cgap:=\pi^2/4$.
  This shift by~$\gamma_L$ scales like the gap between the first
  and the second eigenvalue of~$-\Laplace^L$. 

  The normalized ground state~$\Psi_L$ of~$H_0^L$ is given by
  $\Psi_L=\vol{\Lambda_L}^{-1/2}\ifu{\Lambda_L}$.
  Furthermore, 
  \begin{equation*}
    E_1(H_0^L)
    =-\gamma_L
    \qtextq{and}
    E_2(H_0^L)
      =\frac{\Cgap}{L^2}-\gamma_L
      =\gamma_L
    \quad(L\in\bN)\text.
  \end{equation*}

  As~$V_\omega$ does not vanish,
  $V_\omega^{-1}$ is well-defined as a multiplication operator.
  By construction, we have
  \begin{align*}
    \spr{\Psi_L}{V_\omega^{-1}\Psi_L}&
      =\int \frac{\abs{\Psi_L(x)}^2}{V_\omega(x)}\,\dx
      =\frac1{\vol{\Lambda_L}}
        \int_{\Lambda_L}\frac{\dx}{V_\omega(x)}\\&
      =\frac1{\setsize{I_L}}\sum_{k\in I_L}
        \int_{\cD+k}\frac{\dx}{V_\omega(x)}\text.
  \end{align*}
  The last integral is easily calculated:
  \begin{align*}
        \int_{\cD+k}\frac{\dx}{V_\omega(x)}&
      =
        \int_{\cD}\frac{\dx}{u_{k,\omega}(x)+\gamma_L}\\&
      =\frac{\vol{\lambda_k(\omega)A}}{1+\gamma_L}
            +\frac{1-\vol{\lambda_k(\omega)A}}{\gamma_L}
      =\frac{1+\gamma_L-\vol{\lambda_k(\omega)A}}{(1+\gamma_L)\gamma_L}\text.
  \end{align*}
  With $S_L:=\frac1{\setsize{I_L}}\sum_{k\in I_L}\vol{\lambda_kA}$, we get
  \begin{align*}
    \spr{\Psi_L}{V_\omega^{-1}\Psi_L}&
      =\frac{1+\gamma_L-S_L(\omega)}{(1+\gamma_L)\gamma_L}\text,
  \end{align*}
  or
  \begin{equation*}
    E_1(H_0^L)+\spr{\Psi_L}{V_\omega^{-1}\Psi_L}^{-1}
      =\frac{\gamma_LS_L(\omega)}{1+\gamma_L-S_L(\omega)}\text.
  \end{equation*}
  For all $L\ge L_0:=\sqrt{\Cgap/2}$,
  we have $\gamma_L\le1$.
  Using this as well as $0\le S_L\le1/2$ a.\,s., we derive
  \begin{equation*}
    \frac{\gamma_LS_L(\omega)}2
    \le E_1(H_0^L)+\spr{\Psi_L}{V_\omega^{-1}\Psi_L}^{-1}
    \le\gamma_L
    =E_2(H_0^L)\text.
  \end{equation*}
  Thus, Thirring's inequality implies for all $L\in\bN$, $L\ge L_0$
  \begin{align*}
    E_1(H_\omega^L)&
    =E_1(H_0^L+V_\omega^L)\\&
    \ge\min\{E_1(H_0^L)+\spr{\Psi_L}{V_\omega^{-1}\Psi_L}^{-1},E_2(H_0^L)\}
    \ge\frac{\gamma_LS_L(\omega)}2\text.
  \end{align*}
  From our assumptions on~$A$ and~$\lambda_0$, we have $\E[S_L]=\E[\vol{\lambda_0A}]>0$.
  Let $L_E:=\bigfloor{\sqrt{\Cgap\E[S_1]/(8E)}}$.
  For~$E$ small enough, $L_E\ge L_0$.
  Hence, since~$\E[S_1]=\E[S_{L_E}]$, we see
  \begin{align*}
    \Prob\{\omega\mid E_1(H_\omega^{L_E})\le E\}&
      \le\Prob\bigl\{\tfrac{\gamma_{L_E}}2S_{L_E}\le E\bigr\}
      \le\Prob\{S_{L_E}\le\tfrac12\E[S_{L_E}]\}\text.
  \end{align*}
  Finally, observe that the random variables $\vol{\lambda_kA}$, $k\in\bZ$,
  are independent.
  Bernstein's inequality bounds the last probability by
  $\exp(-\Cld(2L_E)^d)$ with some positive constant~$\Cld$,
  since $\setsize{I_{L_E}}=(2L_E)^d$.
  Restricting $E$ further to be smaller than $C_3 \E[S_1] /32$, 
  we see, from the definition of~$L_E$,
  \begin{equation*}
    N(E)
      \le\Cfront E^{d/2}\exp\bigl(-\Cld(2L_E)^d\bigr)
      \le\Cfront E^{d/2}\exp\bigl(-\Cexp E^{-d/2}\bigr)
  \end{equation*}
  with $\Cexp=\Cld\bigl(\Cgap\E[\vol{\lambda_0A}]/8\bigr)^{d/2}$.
\end{proof}

\subsection*{Acknowledgemts}
This work has been partailly supported by the DFG under grants \emph{Unique continuation principles and
equidistribution properties of eigenfunctions} and 
\emph{Multiscale version of the Logvinenko-Sereda Theorem}.

\def\polhk#1{\setbox0=\hbox{#1}{\ooalign{\hidewidth
  \lower1.5ex\hbox{`}\hidewidth\crcr\unhbox0}}}

\end{document}